\documentclass[seceq,preprint]{ptptex}

\usepackage{graphicx}
\usepackage{amssymb,amsmath}

\newcommand{\figwidth}{0.5\columnwidth}

\title{Tunnelling in organic superconductors}

\author{
C. J. \textsc{Bolech} and T. \textsc{Giamarchi}}

\inst{University of Geneva, 24 Quai Ernest Ansermet, 1211 Geneva,
Switzerland}

\abst{We discuss the possibility of deciding on the symmetry of the
superconducting phase in the organic superconductors (Bechgaard
salts), using tunnelling experiments. We first briefly review the
properties of organic superconductors, and the possibility to have
unconventional (triplet) superconductivity in these systems. We then
present a simple scheme for computing the full current-voltage
characteristics for tunnelling experiments within the framework of the
non-equilibrium Keldysh Green function formalism. This formalism is
flexible enough to address different pairing symmetries combined
with magnetic fields and finite temperatures at arbitrary bias
voltages. We then discuss extensively how to apply these results to
probe for the symmetry of the superconducting order parameter.}

\begin{document}

\maketitle

\section{Introduction}

Organic materials have provided physicists with an extraordinary
laboratory to study the effects of interactions in solids. These
materials indeed show remarkable effects due to the interplay of
interactions and dimensionality\cite{review_organics_complete}.
Indeed such compounds made of coupled chains allow to realize
practically the physics of Luttinger
liquids\cite{giamarchi_book_1d}, one of the very few controlled
examples of non-Fermi liquids\cite{landau_fermiliquid_theory_static}
due to interactions. But contrarily to other realizations of one
dimensional systems such as nanotubes
\cite{dresselhaus_book_fullerenes_nanotubes,bockrath_nanotube_luttinger,yao_nanotube_kink},
quantum
wires\cite{tarucha_quant_cond,auslaender_quantumwire_tunneling,tserkovnyak_quantumwire_tunneling} or
edge states in the quantum Hall
effect\cite{fisher_transport_luttinger_review,glattli_fqhe_review,milliken_edge_states},
the organics offer unique challenges. Indeed, because of their very
three dimensional nature, they provide not a single one dimensional
electron gas, but a very large number of such one dimensional
systems coupled together. This allows thus for a unique new physics
to emerge where the system is able to crossover from a one
dimensional behavior to a more conventional three dimensional one.

As a consequence of this dimensional crossover, at low temperature
these materials undergo instabilities towards three-dimensionally
ordered states, such as spin-Peierls, antiferromagnetic and even
superconducting states. Needless to say, the presence of
superconductivity in these compounds is a tantalizing and challenging
question. Despite a period of quarter of a century since the discovery
of superconductivity in these
materials\cite{jerome_superconductivity_discovery}, the mechanism and
even the symmetry of this superconducting phase have remained elusive,
and many efforts have been devoted to this subject.  Recently, the
case for triplet nature of this superconducting phase has been
made\cite{ishiguro2002}, mostly by measurements of the upper critical
field and by NMR measurements, but the subject is far from being
settled. If true, this would put the organics in the relatively small
club of condensed matter materials where interactions can lead to
unusual triplet superconductivity\cite{sigrist1991}, other examples
being $^{3}$He (a superfluid) and Sr$_{2}$RuO$_{4}$
\cite{rice1995,mackenzie2003}, and can certainly shed light on the
paring mechanism.

In these notes we explain how point contact tunnelling experiments
can be used to probe for the symmetry of a superconducting state. In
the recent years, the possibilities to perform point contact
tunnelling have been drastically enhanced with the development of
scanning tunnelling microscopy (STM)\cite{wolf1989,binnig1999}.
Correspondingly, theories to interpret tunnelling experiments in
superconductors have evolved from simple semiconducting band models
\cite{blonder1982,octavio1983}, to more systematic approaches based
on the tunnelling
Hamiltonian.\cite{cohen1962,wilkins1969,cuevas1996}. We show here
how one can extend and simplify the formalism to make it more
versatile and easy to
implement\cite{bolech_tunnelling_short,bolech_tunnelling_long}. This
allows to study the case of superconducting (singlet or triplet)
leads, as well as the effects of magnetic fields on the junction.
The resulting theory can be thus directly used as a probe of the
symmetry of the leads. We thus discuss how such a probe can be used
for the case of the organics.

The plan of these notes is as follows. In Sec.~\ref{sec:supra} we
briefly recall the salient points of the physics of organic
compounds. In Sec.~\ref{sec:tunnel} we give an outline of our
formalism for the tunnelling between triplet
superconductors\cite{bolech_tunnelling_short,bolech_tunnelling_long}.
Sec.~\ref{sec:test} discusses how to test for triplet
superconductivity in the organics. Finally some conclusions and
perspectives can be found in Sec.~\ref{sec:conclusion}.

\section{A brief word on organics}\label{sec:supra}

Let us briefly recall some of the properties of the organic
materials. We will focuss here on the Bechgaard salts TMTSF$_2$X
which were the first organic compounds to exhibit superconductivity,
and have thus been the focus of intense experimental and theoretical
studies. A general review on these, and parent compounds (so called
Fabre salts TMTTF$_2$X), can be found in
Ref.~\citen{review_organics_complete}.

\begin{figure}
\centerline{\includegraphics[width=\figwidth]{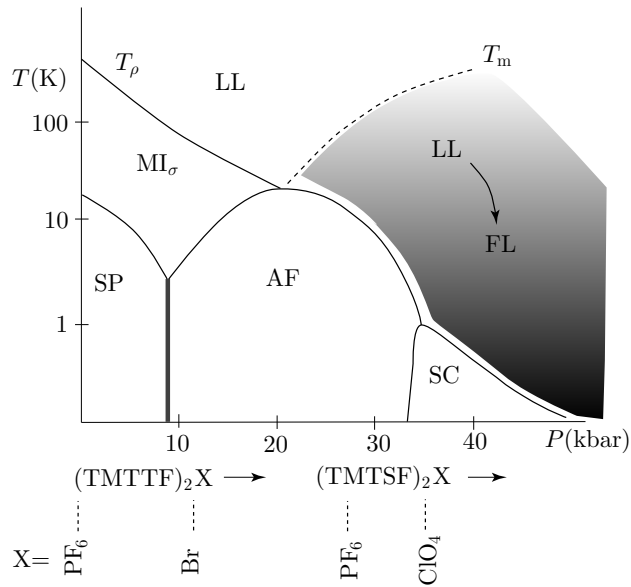}}
 \caption{Unified experimental phase diagram for the TM compounds
   (from Ref.~\citen{bourbonnais_jerome_review}). Either pressure or
   chemical changes (increasing pressure corresponds to either going
   from the TMTTF to the TMTSF family or changing the anions) yields
   the same phases [MI: Mott insulator, LL: Luttinger liquid metal,
   FL: Fermi liquid metal, SP: spin-Peierls, AF: antiferromagnetic
   spin-density wave, SC: superconducting]. Note that the pressure
   axis does not change the doping of the material that remains a
   quarter filled system. The TMTTF family is insulating at ambient
   pressure whereas the TMTSF family shows good metallic behavior at
   room temperature. The superconducting phase, stemming out of the
   antiferromagnetic one is still very poorly understood.}
 \label{fig:unified}
\end{figure}

In addition to the superconducting phase itself, these materials
have a notoriously rich phase diagram (cf. Fig.~\ref{fig:unified})
and exhibit a host of remarkable properties (non-FL metallic
behavior, quantized Hall conductance, Fr{\"o}hlich conductivity),
many of which are still poorly understood.
These materials exhibit a quasi-one dimensional structure due to the
nature of the molecular orbitals protruding from the flat molecule
TMTSF (tetramethyltetraselenafulvalene) which is the basic building
block of the Bechgaard salts: the overlap of the $\pi$- orbitals of
the selenium or sulfur atoms leads to a high mobility of electrons
along the stacking direction. The hopping integrals in the
perpendicular directions are smaller by more than one order of
magnitude. Estimated values of the hopping integrals along the stack
direction (a-axis) and the two perpendicular axes pointing towards
neighboring stacks (b-axis) and towards the anions (c-axis)
respectively are: $t_a:t_b:t_c = 1000K:100K:30K$. Therefore one can
think of these materials as one-dimensional chains coupled by small
inter-chain hoppings. Given the hierarchy of transverse coupling the
system is first expected to become two dimensional and then three
dimensional at low temperatures. The normal phase of these materials
presents unusual transport
\cite{jerome_organic_review,jerome_review_chemrev,dressel_transport_tmtsf}
and optical conductivity properties
\cite{dressel_optical_tmtsf,schwartz_electrodynamics} that have been
interpreted as the signature of Luttinger liquid
physics\cite{giamarchi_mott_shortrev,schwartz_electrodynamics}. The
high temperature phase of these materials thus clearly shows effects
of strong correlations. As the temperature is lowered, the material
recovers features that are more and more reminiscent of a normal
Fermi-liquid material, and finally undergoes instabilities towards
various ordered phases (spin-Peierls (SP), antiferromagnetic (AF),
spin-density wave(SDW)) and superconducting (SC)). The nature of the
molecule (TMTTF vs TMTSF) or the ions, as well as the application of
pressure allows to control the phase diagram, by changing the
hopping integrals. This modifies the relative importance of the
kinetic energy and Coulomb interaction. The chemical and pressure
changes have similar effects, which can be summarized by the unified
phase diagram of Fig.~\ref{fig:unified}. More issues on the effect
of interactions and the properties of the normal phases can be found
in Ref.~\citen{giamarchi_review_chemrev} and references therein.

Among the ordered phases, the superconducting one
\cite{jerome_superconductivity_discovery} is by far the most
mysterious. The mechanism behind it is still unknown, and
considerable debate takes place on that point (see e.g.
Ref.~\citen{ISCOM2003,oh2004,joo2004} and references therein). Since
at the temperature when superconductivity occurs, the compound is
not in the one dimensional limit anymore, but well in the three
dimensional regime, one could in principle think of a conventional
BCS mechanism. However the proximity of the AF phase as well as the
fact, clear from the high temperature phase, that interactions are
particularly strong in these compounds make such a simple pairing
mechanism very unlikely. Besides the mechanism itself, the symmetry
of its superconducting order parameter seems to be spin-triplet
\textit{p}-wave\cite{lebed2000} but claims of \textit{f}-wave
orbital symmetry were also put forward\cite{cherng2003,tanaka2004};
the microscopic origin for the pairing remains largely a matter of
debate. One of the most striking features is the existence of a
common boundary between an antiferromagnetic phase (SDW) and the
superconducting phase that is stabilized as pressure is increased.
Remarkably, the largest values of $T_{c}$ are actually reached near
this boundary, at which the system is expected to display an
enhanced $\mathrm{SO}\left( 4\right)$ symmetry.\cite{podolsky2004}
Intriguingly, recent experiments have identified a pressure window
around the boundary at which the SDW and the superconducting regions
seem to segregate.\cite{vuletic2002} Among the properties of the
superconducting phase that would be in favor of triplet
superconductivity are the sensitivity to non-magnetic
impurities,\cite{joo2004}, the absence of Knight shift signals at
the superconducting transitions\cite{lee2002c} and, perhaps the most
striking, the anomalously large values of the measured upper
critical fields.\cite{lee2002a} Given the nature of these systems,
it is difficult to perform phase sensitive experiments such as the
ones done for the cuprates, so other probes are needed to
unambiguously decide on the symmetry of the order parameter. We now
examine tunnelling as such a probe.

\section{Tunnelling}\label{sec:tunnel}

The earliest successful theoretical models to study superconducting
tunnelling junctions were based on a scattering picture and
semiconducting-like
bands.\cite{nicol1960,klapwijk1982,blonder1982,octavio1983} It was
later shown that those results can be recovered using a tunnelling
Hamiltonian as the starting
point.\cite{cohen1962,wilkins1969,cuevas1996} A large series of
experiments on atomic-size
contacts\cite{scheer1997,scheer1998,ludoph2000,scheer2001,rubio2003,hafner2004}
showed impressive agreement with the theory by considering a small
number of independent conduction channels, each of them well
described by a point-contact model. Tunnelling is thus a very
efficient probe of the properties of the junctions, and one can
expect to use it to determine the symmetry of the superconducting
order parameter in the junction leads. However the techniques used
so far are either semi-phenomenological or very heavy, and one thus
need to find a formulation that is both simple enough to be extended
to the case of unusual superconductors and to finite magnetic fields
and finite temperature, and yet accurate enough to compute the full
current-voltage curve. We describe here such a technique, based on
the Keldysh formalism. More details can be found in
Ref.~\citen{bolech_tunnelling_short,bolech_tunnelling_long}.

Let us use a tunnelling Hamiltonian formalism to calculate the full
current-voltage characteristics of different types of tunnel
junctions where each side of the junction can be either a normal
metal (N), a singlet (S) or a triplet (T) superconductor. We model
the system with a (one-dimensional)
Hamiltonian that includes the two leads and a tunnelling term: $H=H_{1}%
+H_{2}+H_{\mathrm{tun}}$. Each lead is described by%
\begin{equation}
K=\xi_{ck\sigma}\psi_{ck\sigma}^{\dagger}\psi_{ck\sigma}^{%
\phantom{\dagger}%
}-\left\{  \Delta_{a}\left[  \psi_{Rk\beta}^{\dagger}~\sigma_{\beta\alpha}%
^{a}~\alpha~\psi_{L\bar{k}\bar{\alpha}}^{\dagger}\right]  +\mathrm{h.c.}%
\right\}
\end{equation}
where $K=H-\mu N$ and $\mu$ is the corresponding electrochemical potential.
All the indices are summed over, in particular $k$ is the lattice
momentum, Greek indices correspond to the spin, and $c\in\left(
L,R\right)  \equiv\left( -1,+1\right)$ sums over the two possible
chiralities. $\xi_{ck\sigma }=cv_{\mathrm{F}}k-\mu-\sigma h$ are the
corresponding linear dispersions, shifted by the inclusion of
chemical potential and magnetic field along the $\hat{z}$-axis (for
convenience we will take $v_{\mathrm{F}}=1$). This is the extension
of the pairing-approximation Hamiltonian found in BCS theory to the
triplet case (for more details see
Refs.~\citen{bolech_tunnelling_short,bolech_tunnelling_long}). The
third term in the Hamiltonian describes the tunnelling:%
\begin{equation}
H_{\mathrm{tun}}=\sum_{\ell,\ell^{\prime},\sigma}t_{\ell\ell^{\prime}}%
~\psi_{\ell\sigma}^{\dagger}\left(  0\right)
\psi_{\ell^{\prime}\sigma}^{ \phantom{\dagger} }\left(  0\right)
~\text{.}
\end{equation}
According to this term an electron with spin $\sigma$ can hop from
lead $\ell^{\prime}$ into lead $\ell$ with a tunnelling matrix
amplitud given by
\begin{equation}
t_{\ell\ell^{\prime}}=%
\begin{pmatrix}
0 & t^{\ast}\\
t & 0
\end{pmatrix}
~\text{.}%
\end{equation}
Since the number of particles in each lead is a conserved quantity
in the absence of tunnelling (pair fluctuations are to be regarded
as an artifact of the mean field approach), we can define the
current as given by the rate of
change in the relative particle number caused by tunnelling. One writes%
\begin{equation}
I=\frac{e}{2i}\left\langle \left[
H_{\mathrm{tun}},N_{1}-N_{2}\right]
\right\rangle ~\text{.} \label{eq:currentdef}%
\end{equation}

Having in mind that very simple models of the leads suffice to
achieve even quantitative agreement with the experiment when
calculations are carried out to capture the main features of
point-contact transport on conventional superconductors, with
dimensionality playing little or no role, it is justified to use
one-dimensional leads to carry out all the standard calculations.
For unconventional superconductors the situation is more complex,
because the anisotropic nature of the pair wave-function has to be
taken into account when modelling the leads. But the organic
superconductors that we are interested in are supposed to have
\textit{p}-wave symmetry and be highly anisotropic. We can
therefore, as a first approximation, conveniently restrict ourselves
to a one-dimensional model and adopt a formalism that encompasses
both \textit{s}- and \textit{p}-wave symmetries, as well as the
normal state.

In order to deal with an out of equilibrium situation, we use the
so-called Keldysh formalism.\cite{keldysh1965} We treat the
tunnelling term to all orders to calculate the full I-V line and
give a quantitative account of the subgap
structure. Notice that in this framework the current is given by%
\begin{equation}
I=\frac{et}{2i}~\sum_{\sigma}\int\frac{d\omega}{2\pi}\left\langle
\psi_{2,\sigma}^{\dagger}\left(  0\right)  \psi_{1,\sigma}^{
\phantom{\dagger}}\left(  0\right) -\psi_{1,\sigma}^{\dagger}\left(
0\right) \psi_{2,\sigma }^{ \phantom{\dagger}
}\left(  0\right)  \right\rangle _{\mathrm{kel}} \label{eq:curkel}%
\end{equation}
where `\textrm{kel}' denotes the Keldysh component of the
correlation function. Since the current depends only on the fields
at $x=0$ one can
integrate the $x$ dependence in the leads to obtain from $H$ a \textit{local}%
\emph{ }\textit{quadratic} action for the contacts: $S_{\mathrm{jun}}%
=S_{1}+S_{2}+S_{\mathrm{tun}}$. With $S_{\mathrm{tun}}$ obtained
directly from $H_{\mathrm{tun}}$ and
$S_{\ell}=\int\frac{d\omega}{2\pi}\mathbf{\Psi}_{\ell
}^{\dagger}(\omega)\hat{g}^{-1}\mathbf{\Psi}_{\ell}^{
\phantom{\dagger} }(\omega)$. Here $\mathbf{\Psi}_{\ell}$ is an $8$
component spinor and $\hat{g}^{-1}$ is a matrix whose inverse
($\hat{g}$) is given by the standard advanced, retarded and Keldysh
components of the local Green functions of the
lead. For instance, for the case where $\Delta_{1}=\Delta_{2}=0$,%
\begin{align*}
g_{c\sigma,c\sigma}^{\left[  \mathrm{ret,adv}\right]  }  &
=\frac{-\left( \omega-\mu_{\ell}+c\sigma h\pm i0^{+}\right)
}{2\sqrt{\left\vert \Delta _{0}+c\sigma\Delta_{3}\right\vert
^{2}-\left(  \omega-\mu_{\ell}+c\sigma h\pm
i0^{+}\right)  ^{2}}}\\
g_{c\sigma,\bar{c}\bar{\sigma}}^{\left[  \mathrm{ret,adv}\right]  }
& =\frac{\left(  \Delta_{0}+c\sigma\Delta_{3}\right)  ^{\left[
\ast\right] _{c=L}}}{2\sqrt{\left\vert
\Delta_{0}+c\sigma\Delta_{3}\right\vert
^{2}-\left(  \omega-\mu_{\ell}+c\sigma h\pm i0^{+}\right)  ^{2}}}%
\end{align*}
And the Keldysh component is $g^{\mathrm{kel}}=\left(  g^{\mathrm{ret}%
}-g^{\mathrm{adv}}\right)  \tanh\left(  \left(
\omega-\mu_{\ell}\right) /2T\right)$.

\begin{figure}
 \begin{center}
 \includegraphics[width=\figwidth]{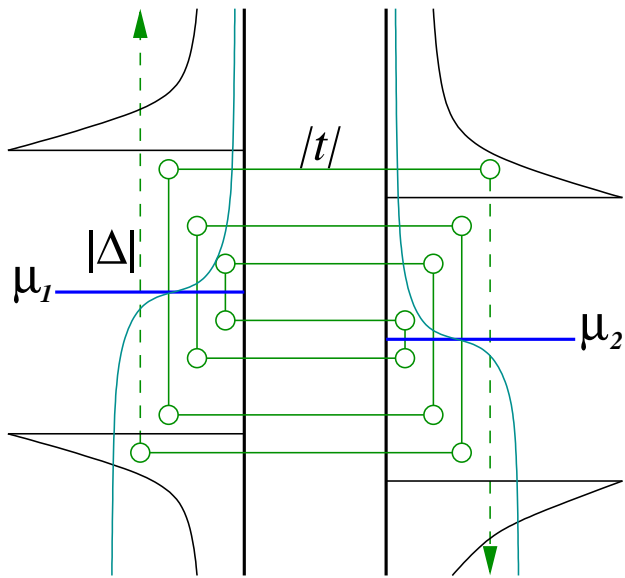}
 \caption{Representation of a set of frequencies involved in a
   high-order tunnelling process (the vertical axis corresponds to
   frequencies). The chemical potentials, thermal distributions and
   superconducting quasiparticle densities of states in each side of
   the junction are all indicated. The figure illustrates the lowest
   order contribution to tunnelling at low voltages for a case that
   corresponds to a process involving one electron plus three Cooper
   pairs.}
 \label{andreev}
 \end{center}
\end{figure}

Attention must be paid to the fact that frequencies have different
reference levels in each side of the junction when a bias is
present. While, in each side of the junction, states of frequencies
with equal positive and negative shifts from the Fermi level are
related by the pairing fluctuations, across the junction, same
frequency states are related by tunnelling (see Fig.~\ref{andreev}).
Thus, the full action for the junction is not diagonal. To each
value ($\omega_{0}$) in the frequency window (of size
$eV=\mu_{1}-\mu_{2}$) defined by the chemical potentials in the two
leads, one infinite set of related frequencies can be assigned
($p>0$):
\begin{equation}\label{q:freq}
\left\{
\begin{array}[c]{r}
\omega_{p}=2\mu_{2-p\operatorname{mod}2}-\omega_{p-1}\\
\omega_{-p}=2\mu_{1+p\operatorname{mod}2}-\omega_{1-p}%
\end{array}
\right.
\end{equation}
The action is block diagonal between these sets. Discretizing the
frequencies in the window defined by the voltage difference defines
in turn a discretization of the whole frequency axis. We shall deal
with one set of frequencies at a time, and since the sets are
infinite, we will truncate their hierarchy. This amounts to
introducing a limit in the number of Andreev reflections. One can then
numerically invert the action block by block (for more details see
Refs.~\citen{bolech_tunnelling_short,bolech_tunnelling_long}). The
off-diagonal Green function matrix elements thus obtained allow to
compute the current using Eq.~(\ref{eq:curkel}).

\section{Test for triplet}\label{sec:test}

As it was stated above, in the Bechgaard salts, both the orbital and
the spin symmetries of the superconducting phase are not clearly
known, not to speak of the microscopic mechanism responsible for the
pairing. Since the uncertainties are such, we think the best way to
proceed would be if the experiments try to resolve these issues one
at a time and to our opinion the most approachable one, from the
point of view of tunnelling, is the issue of the symmetry in spin
space of the superconducting order parameter. Thus, the question to
be answered is whether the electrons in a Cooper pair form a singlet
or a triplet.

To discuss and compare the I-V characteristics for different types
of junctions, we choose some convenient set of parameters that we
will use in all the figures. For the tunnelling overlap integral we
choose the values $t=0.2$ and $t=0.5$ (that correspond, in the
notation of Ref.~\citen{blonder1982}, to $\alpha\simeq0.15$ or
$Z=2.4$ and to $\alpha=0.64$ or $Z=0.75$, respectively), and when
there is a magnetic field we fix its value to $h=0.2$ in units of
$\Delta$ (by $\Delta$ we mean the magnitude of the singlet gap,
$\Delta_{0}$, or of the triplet vector order parameter depending on
the case). We show curves for the dc response in the limit of
vanishing temperatures. The geometry of the junction we consider
corresponds to tunnelling perpendicular to the chains of the
quasi-onedimensional compound, with an orbital order parameter
aligned along them. In this situation no mid-gap states are expected
and, therefore, no midgap features either. We concentrate on what
happens at the conduction edge, in particular the effects of applied
fields. We refer the reader to the literature for some recent
studies that look at the effect of fields on zero bias
anomalies.\cite{tanuma2002}
\begin{figure}
 \begin{center}
 \includegraphics[width=\figwidth]{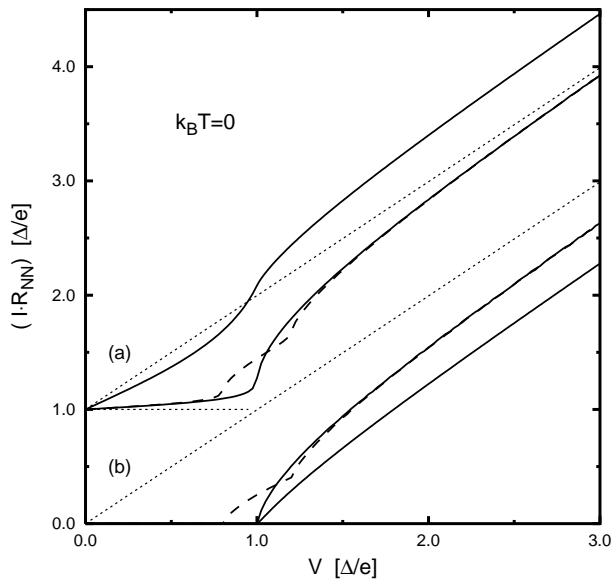}
 \caption{I-V characteristics of
   normal-superconductor junctions. (a) N-S junctions for $t=0.2$
   (lower curves: with and without applied magnetic field, dashed and
   solid line respectively, $h=0.2$) and $t=0.5$ (upper curve, solid
   line only); the curves are vertically displaced for clarity. (b)
   N-T junctions for $t=0.2$ (upper curves) and $t=0.5$ (lower
   curve).}
 \label{nsgraph2}
 \end{center}
\end{figure}

Let us start with the case of normal-metal--superconductor junctions,
that would corespond to standard STM experiments. We show in
Fig.~\ref{nsgraph2}-(a) typical curves for a point-contact junction
between a normal metal and a conventional singlet-paring
superconductor. The solid lines correspond to the N-S junction in zero
field and the dashed line is for one of the junctions in the presence
of a magnetic field that produces what would be seen as a Zeeman
splitting of the differential conductance peak. The second part of the
figure corresponds to a junction between a normal metal and an
unconventional triplet-pairing superconductor. The solid lines
correspond to the N-T junction in zero field and the dashed line is
for the $t=0.2$ junction when in the presence of a magnetic field that
is aligned with the vector order parameter $\vec{\Delta}$. If one
considers a magnetic field perpendicular to the order parameter
($\vec{h}\perp\vec{\Delta}$), it has no effect on the I-V
characteristic.
\begin{figure}
 \begin{center}
 \includegraphics[width=\figwidth]{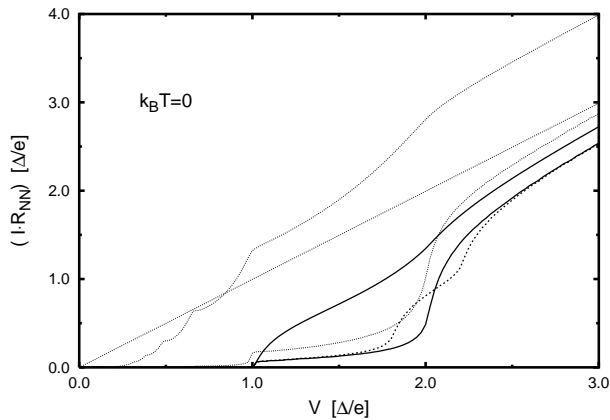}
 \caption{I-V characteristics of S-T junctions with
   and without magnetic field (dashed and solid lines respectively,
   $h=0.2$). The lower solid and dashed curves are for $t=0.2$
   (without and with magnetic field) and the upper solid curve is for
   $t=0.5$.  The dotted lines are (i) the straight unitary slope line
   given for reference and (ii) the S-S characteristics for
   similar-parameters junctions (upper $t=0.5$ and lower $t=0.2$) given
   for comparison purposes.}
 \label{stgraph3}
 \end{center}
\end{figure}
On the other hand, in Fig.~\ref{stgraph3}, we display typical curves
for junctions in which both sides are conventional spin-singlet
superconductors and junctions connecting a spin-singlet and
spin-triplet superconductor. These curves address the situation of a
potential STM experiment carried out using a superconducting tip made
out of a conventional superconductor used to probe a superconducting
phase of unknown symmetry. We remind the reader that, in the case of
conventional superconductors (dotted lines) and when orbital effects
can be ignored, the I-V is not sensitive to applied magnetic fields.
On the other hand, the solid lines correspond to singlet-triplet
junctions, that are insensitive to the orientation of the vector order
parameter on the triple-pairing side of the junctions, and their
current amplitude is found to be systematically smaller than in the
case of the respective singlet-singlet junctions. The `sub-gap'
structure shows only two steps and the current is zero when
$eV<\Delta_{\mathrm{Triplet}}$ (the vector order parameter on the
spin-triplet side of the junction). Concerning the effects of an
applied magnetic field, the curves remain unchanged if the field is
applied parallel to the direction of the vector-order-parameter, but
show instead a Zeeman effect if the field is perpendicular to it
(dashed line).

Regarding the organic
superconductors,\cite{review_organics_complete} the experiments show
that for magnetic fields along the direction of the conducting
chains ($\mathbf{a}$) the upper critical field is possibly
paramagnetically limited for small fields (before crossing the upper
critical field along $\mathbf{b}^{\prime}$ that is never
paramagnetically limited). In that range we could assume that the
direction of the order parameter is fixed respect to the lattice and
does not follow the applied field.\cite{lebed2000} With this
geometry, a Zeeman splitting of the differential conductance peak
should be observed in a normal-tip STM experiment. As the field is
rotated the splitting would be suppressed and for a magnetic field
oriented parallel to the $\mathbf{b}^{\prime}$ crystalline-axis
there should be no Zeeman effect. The disappearance of splitting,
even as the field is possibly being increased, would constitute a
clear signature of spin-triplet superconductivity. Similarly, an
\textit{s}-wave-tip STM would also be a direct probe for
spin-triplet order. When a field is applied along the
$\mathbf{b}^{\prime}$ crystalline-axis, a Zeeman splitting would
occur. This would constitute a clear sign of unconventional
superconductivity since such an effect does not take place for
standard BCS superconductors. The $\mathbf{b}^{\prime}$ direction is
the one on which the upper critical field is not paramagnetically
limited, so relatively large fields could be applied in order to
obtain a clear signal, and as the field alignment changes the
splitting should disappear. No successful attempts of this kind of
experiments were as yet made in the case of the
quasi-one-dimensional organic salts, but efforts in this direction
are on their way (recently, preliminary experiments involving
junctions between two Bechgaard salts were performed, and they
showed a number of puzzling features including a zero-bias
conductance peak and zero excess current).\cite{ha2003} We certainly
hope that further tunnelling experiments along the lines described
on these notes can be performed.

\section{Conclusions and perspectives} \label{sec:conclusion}

In these notes we have presented a new formalism, based on Keldysh
technique, to compute the full current-voltage characteristics of
point contact junctions between normal or superconducting (singlet
or triplet) leads. This formalism is flexible enough to easily
incorporate the effects of magnetic fields or temperature. As a
possible application of this formalism we have shown how it can be
used to devise a tunnelling experiment allowing to probe for the
triplet nature of the superconducting phase in the Bechgaard salts.

Clearly, this formalism can be extended in many ways. One of the most
interesting extensions, in view of recent tunnelling
experiments\cite{ha2003}, is to take into account a finite region for
the tunnelling contact, an intermediate between the point contact and
the planar junction. Other extension, needed for example in the case
of ruthenates, would be to incorporate the modifications of the
superconducting order parameter close to the surface. Such a
difficulty could be avoided for the organics given the large mass
anisotropy, but is more crucial for an isotropic two dimensional
system, since surfaces are pair breaking for triplet superconductors
when the momentum is perpendicular to the surface. Finally, extensions
to other pairing symmetries such as $d$-wave are also possible; such
extensions are, of course, exciting questions to be covered in future
studies.

\section*{Acknowledgements}
We would like to thank Ø. Fischer, M. Eskildsen, M. Kugler, and G.
Levy for discussions about tunneling and STM. We also thank Y. Maeno
and M. Sigrist for discussions about tunneling into triplet
superconductors and ruthenates in particular. Part of this work has
been supported by the Swiss National Science Fundation under MaNEP and
Division II.


\begin{thebibliography}{10}

\bibitem{review_organics_complete}
(2004).
\newblock For recent reviews on organics, see the volume {\bf 104} of Chemical
  Reviews.

\bibitem{giamarchi_book_1d}
T.~Giamarchi.
\newblock {\em Quantum Physics in One Dimension}.
\newblock Oxford University Press, Oxford, (2004).

\bibitem{landau_fermiliquid_theory_static}
L.~D. Landau.
\newblock {\em Sov. Phys. JETP}, {\bf 3} (1957), 920.

\bibitem{dresselhaus_book_fullerenes_nanotubes}
M.~S. Dresselhaus, G.~Dresselhaus, and P.~C. Eklund.
\newblock {\em Science of Fullerenes and Carbon Nanotubes}.
\newblock Academic Press, San Diego, CA, (1995).

\bibitem{bockrath_nanotube_luttinger}
M.~Bockrath, D.H. Cobden, J.~Lu, A.G. Rinzler, R.E. Smalley,
L.~Balents, and
  P.L. McEuen.
\newblock {\em Nature}, {\bf 397} (1999), 598.

\bibitem{yao_nanotube_kink}
Z.~Yao, H.~W.~C. Postma, L.~Balents, and C.~Dekker.
\newblock {\em Nature}, {\bf 402} (1999), 273.

\bibitem{tarucha_quant_cond}
S.~Tarucha, T.~Honda, and T.~Saku.
\newblock {\em Solid State Commun.}, {\bf 94} (1995), 413.

\bibitem{auslaender_quantumwire_tunneling}
O.~M. Auslaender, A.~Yacoby, R.~{de Picciotto}, K.~W. Baldwin, L.~N.
Pfeiffer,
  and K.~W. West.
\newblock {\em Science}, {\bf 295} (2002), 825.

\bibitem{tserkovnyak_quantumwire_tunneling}
Y.~Tserkovnyak, B.~I. Halperin, O.~M. Auslaender, and A.~Yacoby.
\newblock {\em Phys. Rev. Lett.}, {\bf 89} (2002), 136805.

\bibitem{fisher_transport_luttinger_review}
M.~P.~A. Fisher and L.~I. Glazman.
\newblock In L.~{Kowenhoven {\it et al.}}, editor, {\em Mesoscopic Electron
  Transport}, Dordrecht, (1997). Kluwer Academic Publishers.
\newblock cond-mat/9610037.

\bibitem{glattli_fqhe_review}
C.~Glattli.
\newblock In C.~{Berthier {\it et al.}}, editor, {\em High Magnetic Fields:
  Applications in Condensed Matter Physics and Spectroscopy}, page~1, Berlin,
  (2002). Springer-Verlag.

\bibitem{milliken_edge_states}
F.~P. Milliken, C.~P. Umbach, and R.~A. Webb.
\newblock {\em Solid State Commun.}, {\bf 97} (1996), 309.

\bibitem{jerome_superconductivity_discovery}
D.~J{\'e}rome, A.~Mazaud, M.~Ribault, and K.~Bechgaard.
\newblock {\em J. Phys. Lett.}, {\bf 41} (1980), L--95.

\bibitem{ishiguro2002}
T.~Ishiguro.
\newblock Superconductivity under high magnetic fields in low-dimensional
  organic salts.
\newblock In C.~Berthier, L.~P. L{\'e}vy, and G.~Martinez, editors, {\em High
  Magnetic Fields: Applications in Condensed Matter Physics and Spectroscopy},
  volume 595 of {\em Lecture Notes in Physics}, pages 301--313, Heidelberg,
  (2002). Springer-Verlag.

\bibitem{sigrist1991}
M.~Sigrist and K.~Ueda.
\newblock {\em Rev. Mod. Phys.}, {\bf 63} (1991), 239.

\bibitem{rice1995}
T.~M. Rice and M.~Sigrist.
\newblock {\em J. Phys.: Condens. Matter}, {\bf 7} (1995), L643.

\bibitem{mackenzie2003}
A.~P. Mackenzie and Y.~Maeno.
\newblock {\em Rev. Mod. Phys.}, {\bf 75} (2003), 657.

\bibitem{wolf1989}
E.~L. Wolf.
\newblock {\em Principles of Electron Tunneling Spectroscopy}.
\newblock International Series of Monographs on Physics. Oxford Science
  Publications, Clarendon Press, Oxford, second edition, (1989).

\bibitem{binnig1999}
G.~Binnig and H.~Rohrer.
\newblock {\em Rev. Mod. Phys.}, {\bf 71} (1999), S324.

\bibitem{blonder1982}
G.~E. Blonder, M.~Tinkham, and T.~M. Klapwijk.
\newblock {\em Phys. Rev. B}, {\bf 25} (1982), 4515.

\bibitem{octavio1983}
M.~Octavio, M.~Tinkham, G.~E. Blonder, and T.~M. Klapwijk.
\newblock {\em Phys. Rev. B}, {\bf 27} (1983), 6739.

\bibitem{cohen1962}
M.~H. Cohen, L.~M. Falicov, and J.~C. Phillips.
\newblock {\em Phys. Rev. Lett.}, {\bf 8} (1962), 316.

\bibitem{wilkins1969}
J.~W. Wilkins.
\newblock {\em Tunneling Phenomena in Solids}, chapter~24, page 333.
\newblock Multiparticle Tunneling. Plenum Press, New York, first edition,
  (1969).

\bibitem{cuevas1996}
J.~C. Cuevas, A.~Mart{\'{\i}}n-Rodero, and A.~Levy Yeyati.
\newblock {\em Phys. Rev. B}, {\bf 54} (1996), 7366.

\bibitem{bolech_tunnelling_short}
C.~J. Bolech and T.~Giamarchi.
\newblock {\em Phys. Rev. Lett.}, {\bf 92} (2004), 127001.

\bibitem{bolech_tunnelling_long}
C.~J. Bolech and T.~Giamarchi.
\newblock {\em Phys. Rev. B}, {\bf 71} (2005), 024517.

\bibitem{bourbonnais_jerome_review}
C.~Bourbonnais and D.~J{\'e}rome.
\newblock In P.~Bernier, S.~Lefrant, and G.~Bidan, editors, {\em Advances in
  Synthetic Metals, Twenty years of Progress in Science and Technology}, page
  206, New York, (1999). Elsevier.
\newblock preprint cond-mat/9903101.

\bibitem{jerome_organic_review}
D.~J{\'e}rome.
\newblock Organic superconductors: From {(TMTSF)$_2$PF$_6$} to fullerenes.
\newblock page 405, New York, (1994). Marcel Dekker.

\bibitem{jerome_review_chemrev}
D.~J{\'e}rome.
\newblock {\em Chem. Rev.}, {\bf 104} (2004), 5565.

\bibitem{dressel_transport_tmtsf}
M.~Dressel, K.~Petukhov, B.~Salameh, P.~Zornoza, and T.~Giamarchi.
\newblock {\em Phys. Rev. B}, {\bf 71} (2005), 075104.

\bibitem{dressel_optical_tmtsf}
M.~Dressel, A.~Schwartz, G.~Gr{\"u}ner, and L.~Degiorgi.
\newblock {\em Phys. Rev. Lett.}, {\bf 77} (1996), 398.

\bibitem{schwartz_electrodynamics}
A.~Schwartz, M.~Dressel, G.~Gr{\"u}ner, V.~Vescoli, L.~Degiorgi, and
  T.~Giamarchi.
\newblock {\em Phys. Rev. B}, {\bf 58} (1998), 1261.

\bibitem{giamarchi_mott_shortrev}
T.~Giamarchi.
\newblock {\em Physica B}, {\bf 230-232} (1997), 975.

\bibitem{giamarchi_review_chemrev}
T.~Giamarchi.
\newblock {\em Chem. Rev.}, {\bf 104} (2004), 5565.

\bibitem{ISCOM2003}
(2004).
\newblock {ISCOM 2003} - The Fifth International Symposium on Crystalline
  Organic Metals, Superconductors and Ferromagnets. Port-Bourgenay, September
  21{$^{st}$}-26{$^{th}$}, 2003.

\bibitem{oh2004}
J.~I. Oh and M.~J. Naughton.
\newblock {\em Phys. Rev. Lett.}, {\bf 96} (2004), 067001.

\bibitem{joo2004}
N.~Joo, P.~Auban-Senzier, C.~R. Pasquier, P.~Monod, D.~J{\'e}rome,
and
  K.~Bechgaard.
\newblock {\em Eur. Phys. J. B}, {\bf 40} (2004), 43.

\bibitem{lebed2000}
A.~G. Lebed, K.~Machida, and M.~Ozaki.
\newblock {\em Phys. Rev. B}, {\bf 62} (2000), R795.

\bibitem{cherng2003}
R.~W. Cherng and C.~A.~R. {S{\'a} de Melo}.
\newblock {\em Phys. Rev. B}, {\bf 67} (2003), 212505.

\bibitem{tanaka2004}
Y.~Tanaka and K.~Kuroki.
\newblock {\em Phys. Rev. B}, {\bf 70} (2004), 060502(R).

\bibitem{podolsky2004}
D.~Podolsky, E.~Altman, T.~Rostunov, and E.~Demler.
\newblock {\em Phys. Rev. Lett.}, {\bf 93} (2004), 246402.

\bibitem{vuletic2002}
T.~Vuleti{\'c}, P.~{Auban-Senzier}, C.~Pasquier, S.~Tomi{\'c},
D.~J{\'e}rome,
  M.~H{\'e}ritier, and K.~Bechgaard.
\newblock {\em Eur. Phys. J. B}, {\bf 25} (2002), 319.

\bibitem{lee2002c}
I.~J. Lee, S.~E. Brown, W.~G. Clark, M.~J. Strouse, M.~J. Naughton,
W.~Kang,
  and P.~M. Chaikin.
\newblock {\em Phys. Rev. Lett.}, {\bf 88} (2002), 017004.

\bibitem{lee2002a}
I.~J. Lee, P.~M. Chaikin, and M.~J. Naughton.
\newblock {\em Phys. Rev. B}, {\bf 65} (2002), 180502(R).

\bibitem{nicol1960}
J.~Nicol, S.~Shapiro, and P.~H. Smith.
\newblock {\em Phys. Rev. Lett.}, {\bf 5} (1960), 461.

\bibitem{klapwijk1982}
T.~M. Klapwijk, G.~E. Blonder, and M.~Tinkham.
\newblock {\em Physica B \& C}, {\bf 109-110} (1982), 1657.

\bibitem{scheer1997}
E.~Scheer, P.~Joyez, D.~Esteve, C.~Urbina, and M.~H. Devoret.
\newblock {\em Phys. Rev. Lett.}, {\bf 78} (1997), 3535.

\bibitem{scheer1998}
E.~Scheer, N.~Agra{\"{\i}}t, J.~C. Cuevas, A.~Levy Yeyati,
B.~Ludoph,
  A.~Mart{\'{\i}}n-Rodero, G.~Rubio-Bollinger, J.~M. van Ruitenbeek, and
  C.~Urbina.
\newblock {\em Nature}, {\bf 394} (1998), 154.

\bibitem{ludoph2000}
B.~Ludoph, N.~van~der Post, E.~N. Bratus', E.~V. Bezuglyi, V.~S.
Shumeiko,
  G.~Wendin, and J.~M. van Ruitenbeek.
\newblock {\em Phys. Rev. B}, {\bf 61} (2000), 8561.

\bibitem{scheer2001}
E.~Scheer, W.~Belzig, Y.~Naveh, M.~H. Devoret, D.~Esteve, and
C.~Urbina.
\newblock {\em Phys. Rev. Lett.}, {\bf 86} (2001), 284.

\bibitem{rubio2003}
G.~Rubio-Bollinger, C.~de~las Heras, E.~Bascones, N.~Agra{\"{\i}}t,
F.~Guinea,
  and S.~Vieira.
\newblock {\em Phys. Rev. B}, {\bf 67} (2003), 121407R.

\bibitem{hafner2004}
M.~H{\"a}fner, P.~Konrad, F.~Pauly, J.~C. Cuevas, and E.~Scheer.
\newblock Conduction channels of one-atom zinc contacts, (2004).
\newblock cond-mat/0407207.

\bibitem{keldysh1965}
L.~V. Keldysh.
\newblock {\em Sov. Phys. JETP}, {\bf 24} (1965), 1018.

\bibitem{tanuma2002}
Y.~Tanuma, K.~Kuroki, Y.~Tanaka, R.~Arita, S.~Kashiwaya, and
H.~Aoki.
\newblock {\em Phys. Rev. B}, {\bf 66} (2002), 094507.

\bibitem{ha2003}
H.~I. Ha, J.~I. Oh, J.~Moser, and M.~J. Naughton.
\newblock {\em Synth. Metal}, {\bf 137} (2003), 1215.

\end{thebibliography}

\end{document}